%% file: template.tex
\documentclass{Interspeech}

\interspeechcameraready

\title{ASR-FAIRBENCH: Measuring and Benchmarking Equity Across Speech Recognition Systems}

\author{Anand}{Rai}
\author{Satyam}{Rahangdale}
\author{Utkarsh}{Anand}
\author{Animesh}{Mukherjee}

\affiliation{}{Indian Institute of Technology, Kharagpur} {India}
\email{raianand.1991@gmail.com, satyamrahangdale196@kgpian.iitkgp.ac.in, ua28012006@kgpian.iitkgp.ac.in, animeshm@cse.iitkgp.ac.in}
\keywords{speech recognition, fairness benchmark, leaderboard}

\usepackage{comment}
\usepackage{subcaption}

\begin{document}

\maketitle

% the abstract here must exactly match the abstract entered into the paper submission system
\begin{abstract}
    
    % 1000 characters. ASCII characters only. No citations.
  Automatic Speech Recognition (ASR) systems have become ubiquitous in everyday applications, yet significant disparities in performance across diverse demographic groups persist. In this work, we introduce the ASR-FAIRBENCH leaderboard which is designed to assess both the accuracy and equity of ASR models in real-time. 
  
  Leveraging the Meta's  Fair-Speech dataset, which captures diverse demographic characteristics, we employ a mixed-effects Poisson regression model to derive an overall fairness score. This score is integrated with traditional metrics like Word Error Rate (WER) to compute the Fairness Adjusted ASR Score (FAAS), providing a comprehensive evaluation framework. Our approach reveals significant performance disparities in SOTA ASR models across demographic groups and offers a benchmark to drive the development of more inclusive ASR technologies.

\end{abstract}

\input{Introduction}

\input{Methodology}
\input{Application}

\input{Conclusion}

\bibliographystyle{IEEEtran}
\bibliography{mybib}

\end{document}

%% file: Introduction.tex
\section{Introduction}
Automatic Speech Recognition (ASR) systems have revolutionized human-computer interaction, powering applications from virtual assistants to real-time transcription services. However, despite impressive strides in overall accuracy, these systems often exhibit significant performance disparities across diverse demographic groups. Variations in accent, gender, age, and linguistic background can lead to unequal ASR performance ~\cite{tatman2017effects,feng2021quantifying,rai2023deep}, raising concerns about bias and fairness in deployed technologies.

While traditional ASR leaderboards~\cite{wang2024open,open-asr-leaderboard} focus solely on overall accuracy—typically measured by Word Error Rate (WER)—this approach overlooks critical fairness concerns. Many commercial and open-source ASR models are benchmarked on public datasets that lack diverse speech characteristics, perpetuating biases against underrepresented groups~\cite{koenecke2020racial}. As a result, models optimized solely for low WER may perform inadequately for speakers with varied accents, dialects, and other demographic features. Establishing a dedicated fairness leaderboard is essential to ensure that ASR systems are evaluated not only for accuracy but also for equity, promoting more inclusive and responsible technology development.

Our proposed \textbf{ASR-FAIRBENCH} leaderboard utilizes the  Fair-Speech dataset~\cite{veliche2024towards}, a public corpus containing ~26,500 utterances from 593 compensated U.S. participants with self-reported demographics including age, gender, ethnicity, location, and native language status. The dataset spans seven voice assistant domains: music, capture, utilities, notification control, messaging, calling, and dictation.

\textbf{ASR-FAIRBENCH} is the first real-time benchmarking platform evaluating both fairness and accuracy, using 10\% stratified samples from the Fair-Speech dataset. It enhances mixed-effects Poisson regression by integrating it with WER to produce a Fairness-Adjusted ASR Score (FAAS), benchmarks five models, and introduces a five-tier classification scheme from "severely biased" to "exemplarily fair" for overall and attribute-specific assessments.

%% file: Methodology.tex
\section{Methodology}
\subsection{Dataset}
A stratified 10\% sample was extracted from the fair speech dataset to benchmark ASR models on the \textbf{ASR-FAIRBENCH} leaderboard. As shown in Table \ref{tab:dataset_comparison}, the entropy values across all demographic attributes remain nearly identical between the original and sampled datasets, ensuring that the distribution is preserved while reducing the evaluation data by 90\% and significantly decreasing inference time.
\begin{table}[h]
    \centering
    \footnotesize
    \setlength{\abovecaptionskip}{5pt}
    \setlength{\belowcaptionskip}{0pt}
    \begin{tabular}{lcc}
        \toprule
        \textbf{Attributes} & \textbf{Original} & \textbf{Sampled} \\
        \midrule
        Total samples & 26,471 & 2,648 \\
        Total Duration (hrs) & 54.56 & 5.46 \\
        Entropy (Gender) & 0.99 & 0.99 \\
        Entropy (First Lang.) & 1.40 & 1.39 \\
        Entropy (Socioec. Bkg.) & 1.27 & 1.27 \\
        Entropy (Ethnicity) & 2.50 & 2.50 \\
        \bottomrule
    \end{tabular}
    \caption{Original vs. sampled dataset comparison}
    \label{tab:dataset_comparison}
\end{table}
\vspace{-0.8 cm}
\subsection{Evaluation metric} We present a framework for quantifying fairness in ASR systems through the Fairness-Adjusted ASR Score (FAAS).
\vspace{-0.1cm}
\subsubsection{Word error rate and fairness model}
The WER metric evaluates ASR performance: $WER = \frac{S + D + I}{N}$, where $S$, $D$, and $I$ represent substitutions, deletions, and insertions.
To assess fairness across demographic groups, we employ mixed-effects Poisson regression:
\begin{equation}
\log(WER_i) = \beta_0 + \beta_1 X_{i} + \beta_2 Z_{i} + u_i
\end{equation}
where $X_{i}$ represents demographic attributes, $Z_{i}$ represents covariates, and $\beta_1$ quantifies disparity through $\text{disparity ratio} = e^{\beta_1}$.
\vspace{-0.1cm}
\subsubsection{Fairness score calculation}
For each demographic group, we compute predicted WER:
\begin{equation}
\widehat{WER}_g = \frac{e^{(\beta_0 + \beta_g + \beta_{logRef} \cdot \bar{X})}}{e^{\bar{X}} - 1}
\end{equation}
Raw fairness scores are scaled to 0-100:
\begin{equation}
\text{Raw fairness score}_g = 100 \times \left(1 - \frac{\widehat{WER}_g - \min(\widehat{WER})}{\max(\widehat{WER}) - \min(\widehat{WER})}\right)
\end{equation}
The category score is calculated as: $\text{Category score} = \sum_{g} p_g \times \text{Raw fairness score}_g$
\vspace{-0.1cm}
\subsubsection{Statistical adjustment and FAAS}
Statistical significance of fairness disparities is assessed via the Likelihood Ratio Test comparing full and reduced models:
\begin{equation}
LRT = 2 \times (\log L_{full} - \log L_{reduced})
\end{equation}
where $L_{full}$ includes the demographic attribute and $L_{reduced}$ excludes it. The resulting $p$-value indicates whether disparities are statistically significant.
When disparities are significant ($p < 0.05$), we apply a proportional penalty to the category score:
\begin{equation}
\text{Adjusted score} = \text{Category score} \times \left(\frac{p}{0.05}\right)
\end{equation}
For non-significant disparities ($p \geq 0.05$), the category score remains unchanged.
The overall fairness score is calculated as a weighted average across all demographic categories:
\begin{equation}
\text{Overall score} = \frac{\sum_{c} w_c \times \text{Adjusted score}_c}{\sum_{c} w_c}
\end{equation}
where $w_c$ represents optional weighting factors for each category $c$.
Finally, the \textbf{Fairness-Adjusted ASR Score (FAAS)} integrates recognition accuracy with fairness in a single metric:
\begin{equation}
FAAS = 10 \times \log_{10} \left( \frac{\text{Overall Score}}{WER} \right)
\end{equation}
This logarithmic formulation ensures that improvements in both fairness and accuracy contribute positively to the final score, providing a comprehensive evaluation of ASR system performance across diverse demographic groups. The detailed formulation of the \textbf{FAAS} metric can be found in the HuggingFace Space\footnote{\label{myfootnote}{https://huggingface.co/spaces/satyamr196/ASR-FairBench}}.

%% file: Application.tex
\section{Application overview}

\begin{figure}
    \centering
    \begin{subfigure}{0.48\textwidth}
        \centering
        \includegraphics[width=\linewidth, height=4.8cm, keepaspectratio]{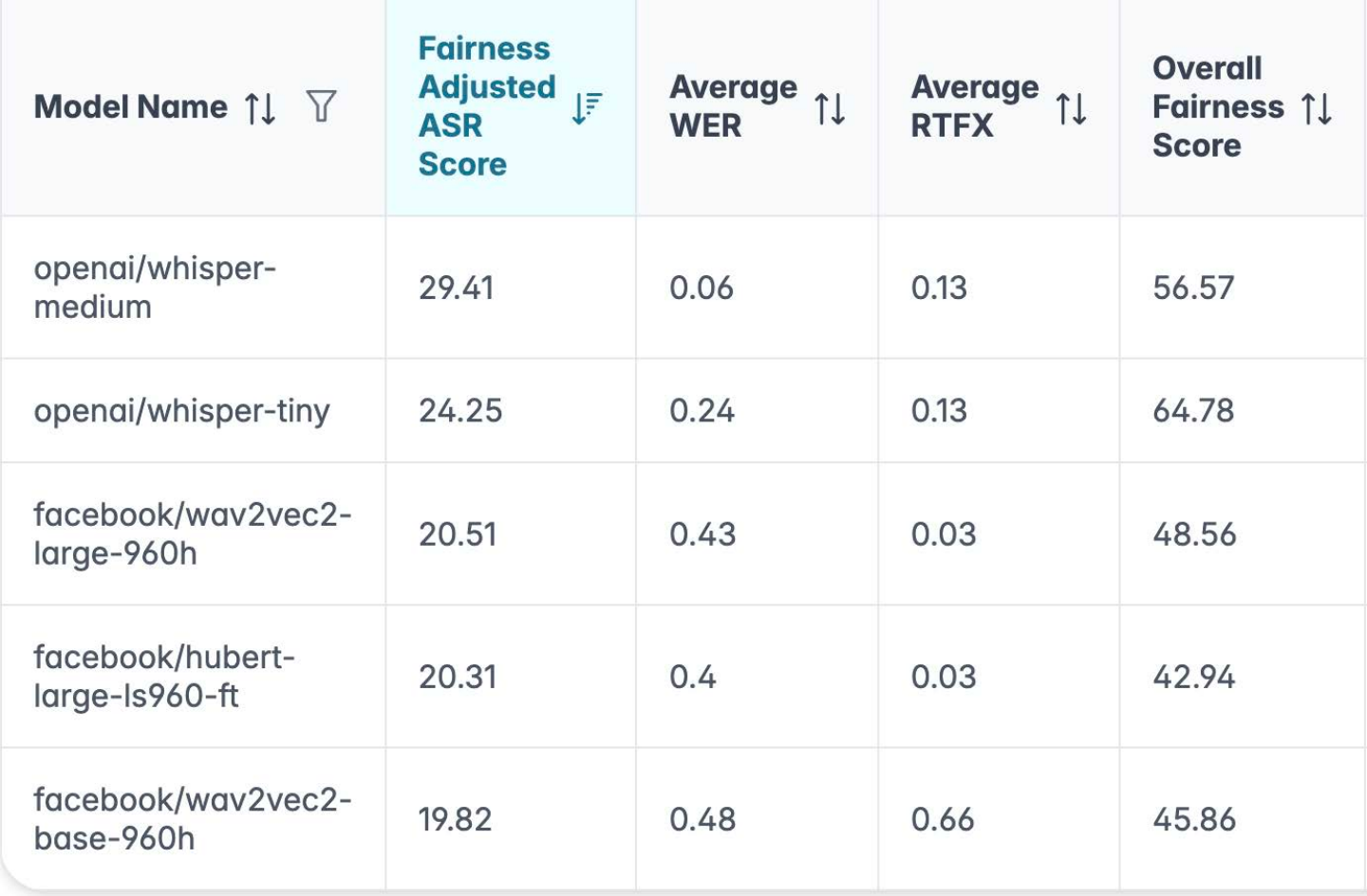}
        \caption{\textbf{ASR-FAIRBENCH} leaderboard}
        \label{fig:submit_model}
    \end{subfigure}
    \hfill
    \begin{subfigure}{0.48\textwidth}
        \centering
        \includegraphics[width=\linewidth]{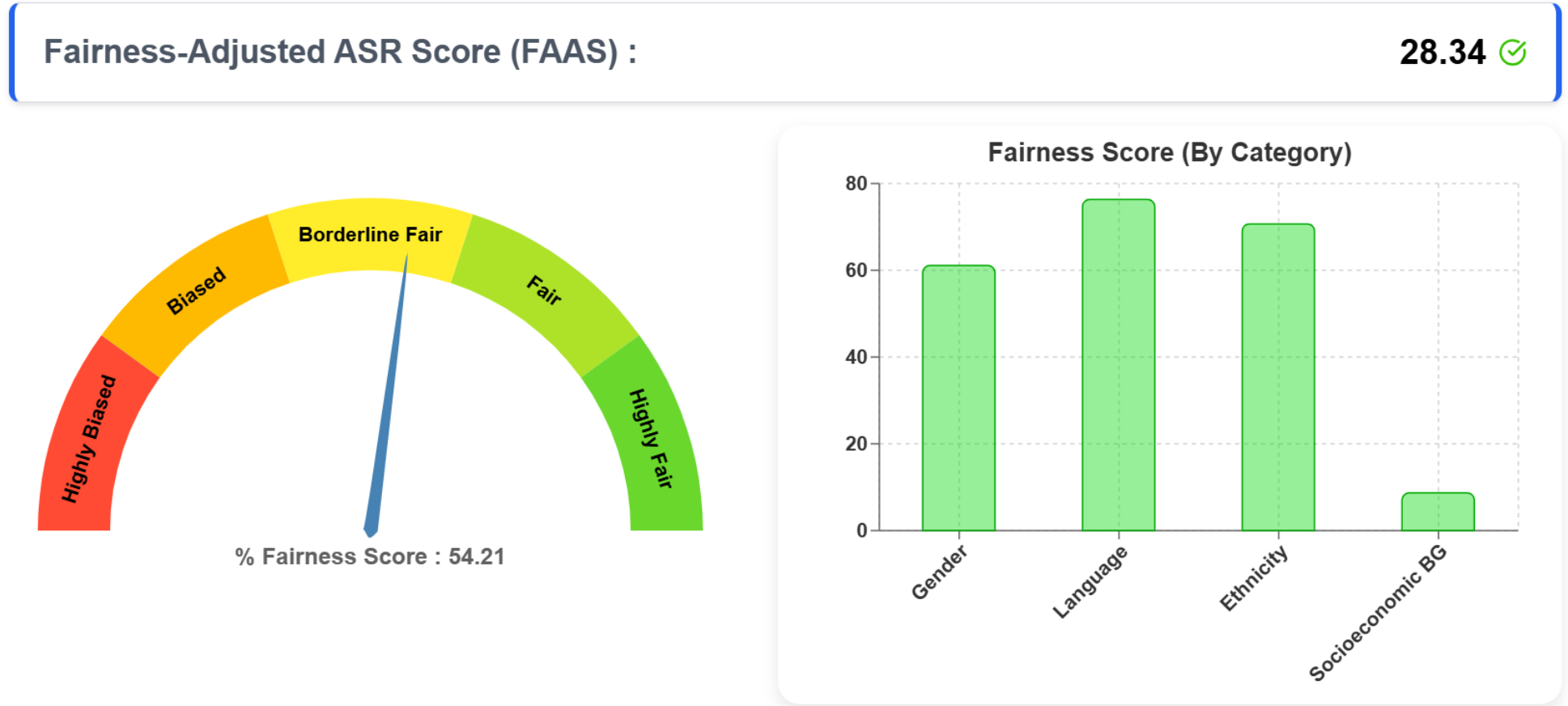}
        \caption{Summarized results view.}
        \label{fig:summary_results}
    \end{subfigure}
    \caption{Key UI features: \textbf{ASR-FAIRBENCH} leaderboard, result summary, and performance analysis for whisper-small.}
    \label{fig:ui_features}
\end{figure}

The \textbf{ASR-FAIRBENCH} leaderboard is a web-based platform designed for evaluating and ranking ASR models. Built with React.js, leveraging an NVIDIA T4 GPU for inference, it offers an interactive platform for model submissions, performance analysis, and real-time leaderboard tracking. The platform is fully reproducible from its  repository\footnote{https://github.com/SatyamR196/ASR-FairBench}. The live leaderboard\footnotemark[\value{footnote}] can be accessed by users to submit their ASR models and receive an audit within minutes. The platform offers intuitive graphical representations of results, including box plots and histograms as shown in Figure \ref{fig:ui_features}, making it easier to interpret model performance.

% \begin{table} 
%     \centering
%     \small
%     \renewcommand{\arraystretch}{1.2}
%     \resizebox{\linewidth}{!}{ % Resize table to fit if needed
%     \begin{tabular}{|l|c|c|c|c|c|c|c|c|}
%         \hline
%         \textbf{Model Name} & \textbf{ASR Score} & \textbf{WER} & \textbf{RTFX} & \textbf{Fairness Score} & \textbf{Gender} & \textbf{First Lang.} & \textbf{Socioecon.} & \textbf{Ethnicity} \\
%         \hline
%         openai/whisper-medium & 29.41 & 0.06 & 0.13 & 56.57 & 59.42 & 90.53 & 12.16 & 64.16 \\
%         openai/whisper-tiny   & 24.25 & 0.24 & 0.13 & 64.78 & 44.05 & 92.68 & 40.14 & 82.23 \\
%         facebook/wav2vec2-large-960h & 20.51 & 0.43 & 0.03 & 48.56 & 54.49 & 70.89 & 20.31 & 48.53 \\
%         facebook/hubert-large-ls960-ft & 20.31 & 0.40 & 0.03 & 42.94 & 12.75 & 88.91 & 14.92 & 55.18 \\
%         facebook/wav2vec2-base-960h & 19.82 & 0.48 & 0.66 & 45.86 & 14.58 & 77.01 & 31.06 & 60.77 \\
%         \hline
%     \end{tabular}}
%     \caption{Comparison of ASR Models Based on Performance and Fairness Metrics}
%     \label{tab:asr_comparison}
% \end{table}

%% file: Conclusion.tex
\section{Conclusion}

Our \textbf{ASR-FAIRBENCH} leaderboard (Figure \ref{fig:submit_model}) highlights Whisper models as leading in fairness-adjusted ASR scores, with Whisper-medium scoring highest at 29.41. Despite higher WER, Whisper-tiny demonstrates better overall fairness. Fine-tuned Wav2Vec and Hubert models show efficiency but lag in fairness. While \textbf{FAAS} generally follows WER trends, the comparison between fine-tuned Wav2Vec-large and Hubert-large reveals that fairness issues can outweigh slight WER improvements—emphasizing the leaderboard’s focus on multidimensional fairness evaluation.